\documentclass[%
 reprint,
 bibnotes,
 amsmath,amssymb,
 aps,
 prl,
 ]{revtex4-1}

\usepackage{graphicx}
\usepackage{dcolumn}
\usepackage{bm}
\usepackage[pdfstartview=FitH,
CJKbookmarks=true,
bookmarksnumbered=true,
bookmarksopen=true,
colorlinks,
linkcolor=blue,
anchorcolor=blue,
citecolor=blue,
allcolors=blue,
pdfborder=001,
]{hyperref}
\usepackage[caption=false]{subfig}
\usepackage{floatrow}
\usepackage{enumitem}
\DeclareSubrefFormat{myparens}{#1~(#2)}
\DeclareCaptionListOfFormat{myparens}{#1(#2)}
\newcommand\T{\rule{0pt}{2.6ex}}       
\newcommand\B{\rule[-1.2ex]{0pt}{0pt}} 

\begin{document}

\title{\boldmath Observation of a state $X(2600)$ in the $\pi^{+}\pi^{-}\eta'$ system in the process $J/\psi\rightarrow\gamma\pi^{+}\pi^{-}\eta'$}

\author{
\begin{small}
\begin{center}
M.~Ablikim$^{1}$, M.~N.~Achasov$^{10,b}$, P.~Adlarson$^{68}$, S. ~Ahmed$^{14}$, M.~Albrecht$^{4}$, R.~Aliberti$^{28}$, A.~Amoroso$^{67A,67C}$, M.~R.~An$^{32}$, Q.~An$^{64,50}$, X.~H.~Bai$^{58}$
% ...
\end{center}
\end{small}
\begin{small}
\begin{center}
\vspace{0.2cm}
(BESIII Collaboration)\\
\vspace{0.2cm} {\it
$^{1}$ Institute of High Energy Physics, Beijing 100049, People's Republic of China\\
% ...
$^{4}$ Bochum Ruhr-University, D-44780 Bochum, Germany\\
% ...
$^{10}$ G.I. Budker Institute of Nuclear Physics SB RAS (BINP), Novosibirsk 630090, Russia\\
% ...
$^{14}$ Helmholtz Institute Mainz, Staudinger Weg 18, D-55099 Mainz, Germany\\
% ...
$^{28}$ Johannes Gutenberg University of Mainz, Johann-Joachim-Becher-Weg 45, D-55099 Mainz, Germany\\
% ...
$^{32}$ Liaoning Normal University, Dalian 116029, People's Republic of China\\
% ...
$^{50}$ State Key Laboratory of Particle Detection and Electronics, Beijing 100049, Hefei 230026, People's Republic of China\\
% ...
$^{58}$ University of Jinan, Jinan 250022, People's Republic of China\\
% ...
$^{64}$ University of Science and Technology of China, Hefei 230026, People's Republic of China\\
% ...
$^{67}$ University of Turin and INFN, (A)University of Turin, I-10125, Turin, Italy; (B)University of Eastern Piedmont, I-15121, Alessandria, Italy; (C)INFN, I-10125, Turin, Italy\\
$^{68}$ Uppsala University, Box 516, SE-75120 Uppsala, Sweden\\
% ...
$^{b}$ Also at the Novosibirsk State University, Novosibirsk, 630090, Russia\\
% ...
}
\end{center}
\vspace{0.4cm}
\end{small}
}

\begin{abstract} 

Based on $(10087\pm44)\times10^{6}$ $J/\psi$ events collected with the
BESIII detector, the process
$J/\psi\rightarrow\gamma\pi^{+}\pi^{-}\eta'$ is studied using two
dominant decay channels of the $\eta'$ meson,
$\eta'\rightarrow\gamma\pi^{+}\pi^{-}$ and
$\eta'\rightarrow\eta\pi^{+}\pi^{-},\eta\rightarrow\gamma\gamma$. The
$X(2600)$ is observed with a statistical significance larger than
20$\sigma$ in the $\pi^{+}\pi^{-}\eta'$ invariant mass spectrum, and
it has a strong correlation to a structure around
1.5$~\mathrm{GeV}$/{\it c}$^{2}$ in the $\pi^{+}\pi^{-}$ invariant
mass spectrum. A simultaneous fit on the $\pi^{+}\pi^{-}\eta'$ and
$\pi^{+}\pi^{-}$ invariant mass spectra with the two $\eta'$ decay
modes indicates that the mass and width of the $X(2600)$ state are
$2617.8\pm2.1 ^{+18.2}_{-1.9}~\mathrm{MeV}$/{\it c}$^{2}$ and $200\pm8
^{+20}_{-17}~\mathrm{MeV}$, respectively.  The corresponding branching
fractions are measured to be $B(J/\psi\rightarrow\gamma X(2600))\cdot
B(X(2600)\rightarrow f_{0}(1500)\eta')\cdot
B(f_{0}(1500)\rightarrow\pi^{+}\pi^{-})$ =
($3.39\pm0.18^{+0.91}_{-0.66})\times10^{-5}$ and
$B(J/\psi\rightarrow\gamma X(2600))\cdot B(X(2600)\rightarrow
f_{2}^{'}(1525)\eta')\cdot
B(f_{2}^{'}(1525)\rightarrow\pi^{+}\pi^{-})$ =
($2.43\pm0.13^{+0.31}_{-1.11})\times10^{-5}$, where
the first uncertainties are statistical, and the second systematic.

\end{abstract}

\maketitle

Radiative decays of the $J/\psi$ meson are ideally suited for light
hadron spectroscopy studies, including in particular searches for exotic
hadrons, e.g. of glueballs and hybrids \cite{PhysRevD.50.3268,
PhysRevD.55.5749}.  The $\pi^{+}\pi^{-}\eta'$ mode is of special
interest since it is one of the most favorite modes to search for a
pseudoscalar glueball \cite{AMSLER200461, Eshraim:2012jv}. Lattice Quantum Chromodynamics(LQCD)
predicts that the ground state of the $0^{-+}$ glueballs has a mass
around 2.3 - 2.6$~\mathrm{GeV}$/{\it c}$^{2}$
\cite{LQCD1,LQCD2,LQCD3,LQCD4}. Therefore, it is important to
search for all possible mesons in the 2.3 -
2.6$~\mathrm{GeV}$/{\it c}$^{2}$ region in
$J/\psi\rightarrow\gamma\pi^{+}\pi^{-}\eta'$ decays with the
unprecedented sample of $J/\psi$ events collected at BESIII.

In the process $J/\psi\rightarrow\gamma\pi^{+}\pi^{-}\eta'$, a set of
exotic states have been observed. The $X(1835)$ resonance was first
observed by the BES collaboration \cite{X(1835)} with a statistical
significance of $7.7\sigma$ and confirmed with a statistical
significance larger than $20\sigma$ by the BESIII collaboration
\cite{X(1835)_confirmed}.  The $X(2120)$ and $X(2370)$ resonances were
first observed with statistical significances of $7.2\sigma$ and
$6.4\sigma$, respectively, by the BESIII collaboration
\cite{X(1835)_confirmed}. The spin parity of the $X(1835)$ resonance
is determined to be $0^{-+}$ in the process $J/\psi\rightarrow\gamma
K_{S}^{0}K_{S}^{0}\eta$ \cite{jpc1835}. The theoretical interpretation
of $X(1835)$ is still in question. Possibilities include a
$p\bar{p}$ bound state \cite{ppbound}, a second radial excitation of
the $\eta'$ \cite{PhysRevD.73.014023} and a pseudoscalar glueball
\cite{KOCHELEV2006283}. The measured mass of $X(2370)$ is consistent
with the LQCD prediction for the pseudoscalar glueball
\cite{LQCD3}. Further observations of the process
$J/\psi\rightarrow\gamma\pi^{+}\pi^{-}\eta'$ are important to help the
understanding of QCD and hadronic physics.

In this Letter, the process
$J/\psi\rightarrow\gamma\pi^{+}\pi^{-}\eta'$ is studied with two
dominant $\eta'$ decay modes, $\eta'\rightarrow\gamma\pi^{+}\pi^{-}$
and $\eta'\rightarrow\pi^{+}\pi^{-}\eta,\ \eta\rightarrow\gamma\gamma$
using the $J/\psi$ data sample collected with the BESIII
detector 
in the years of 2009, 2012, 2018 and 2019. The number of
$J/\psi$ decays is $(10087\pm44)\times 10^{6}$ \cite{number}.  A
resonance, the $X(2600)$, is observed in the $\pi^{+}\pi^{-}\eta'$
invariant mass spectrum.

The BESIII detector \cite{detector} records symmetric $e^{+}e^{-}$
collisions provided by the BEPCII storage ring \cite{store_ring},
which operates with a peak luminosity of $1\times10^{33}$
\textrm{cm}$^{-2}s^{-1}$ in the center-of-mass energy range from 2.0
to 4.7~GeV. The cylindrical core of the
BESIII detector covers $93\%$ of the full solid angle and consists of
a helium-based multilayer drift chamber (MDC), a plastic scintillator
time-of-flight system (TOF), and a CsI(Tl) electromagnetic calorimeter
(EMC), which are all enclosed in a superconducting solenoidal magnet
providing a 1.0 T (0.9 T in 2012) magnetic field. The solenoid is
supported by an octagonal flux-return yoke with resistive plate
counter muon identification modules interleaved with steel. The
charged-particle momentum resolution at 1$~\mathrm{GeV}$/$c$ is
$0.5\%$, and the $dE/dx$ resolution is $6\%$ for electrons from Bhabha
scattering. The EMC measures photon energies with a resolution of
$2.5\%\ (5\%)$ at 1$~\mathrm{GeV}$ in the barrel (end cap) region. The
time resolution in the TOF barrel region is 68~ps, while that in the
end cap region is 110~ps. The end cap TOF system was upgraded in 2015
using multi-gap resistive plate chamber technology, providing a time
resolution of 60~ps \cite{etof1,etof2,etof3}.

Simulated data samples produced with
{\footnotesize{\sc{GEANT4}}}-based \cite{AGOSTINELLI2003250} Monte
Carlo (MC) software, which includes the geometric description of the
BESIII detector and the detector response, are used to determine
detection efficiencies and to estimate backgrounds. The simulation
models the beam energy spread and initial state radiation in the
$e^{+}e^{-}$ annihilations with the generator
{\footnotesize{\sc{KKMC}}} \cite{KKMC,KKMC2}.  An inclusive MC sample
includes both the production of the $J/\psi$ resonance and the
continuum processes incorporated in {\footnotesize{\sc{KKMC}}}
\cite{KKMC,KKMC2}.  The known decay modes are modeled with
{\footnotesize{\sc{EVTGEN}}} \cite{EVTGEN,EVTGEN2} using branching
fractions taken from the Particle Data Group (PDG) \cite{pdg}, and the
remaining unknown charmonium decays are modeled with
{\footnotesize{\sc{LUNDCHARM}}} \cite{Lund-Charm,Lund-Charm2}. Final
state radiation from charged final state particles is
incorporated using {\footnotesize{\sc{PHOTOS}}} \cite{photos}.

Charged tracks detected in the MDC are required to be within the polar
angle ($\theta$) range of $|\!\cos\theta|<0.93$, where $\theta$ is
defined with respect to the $z$-axis, which is the symmetry axis of
the MDC.  The distance of closest approach to the interaction point 
must be less than 10 \textrm{cm} along the z-axis and
less than 1 \textrm{cm} in the transverse plane.  Particle
identification (PID) for charged tracks combines measurements of the
$dE/dx$ in the MDC and the flight time in the TOF to form likelihoods
for each hadron ($p$, $K$ and $\pi$) hypothesis; each track is
assigned to the particle type that corresponds to the hypothesis with
the highest confidence level.  Photon candidates are identified using
showers in the EMC. The deposited energy of each shower must be more
than 25$~\mathrm{MeV}$ in the barrel region
($|\!\cos\theta|<0.8$) and more than 50$~\mathrm{MeV}$ in the
end cap region ($0.86<|\!\cos\theta|<0.92$).  To suppress electronic
noise and showers unrelated to the event, the difference between the
EMC time and the event start time is required to be within (0, 700)
ns.

\begin{figure}[htbp]
	\centering
	\subfloat{
		\includegraphics[width=0.5\textwidth]{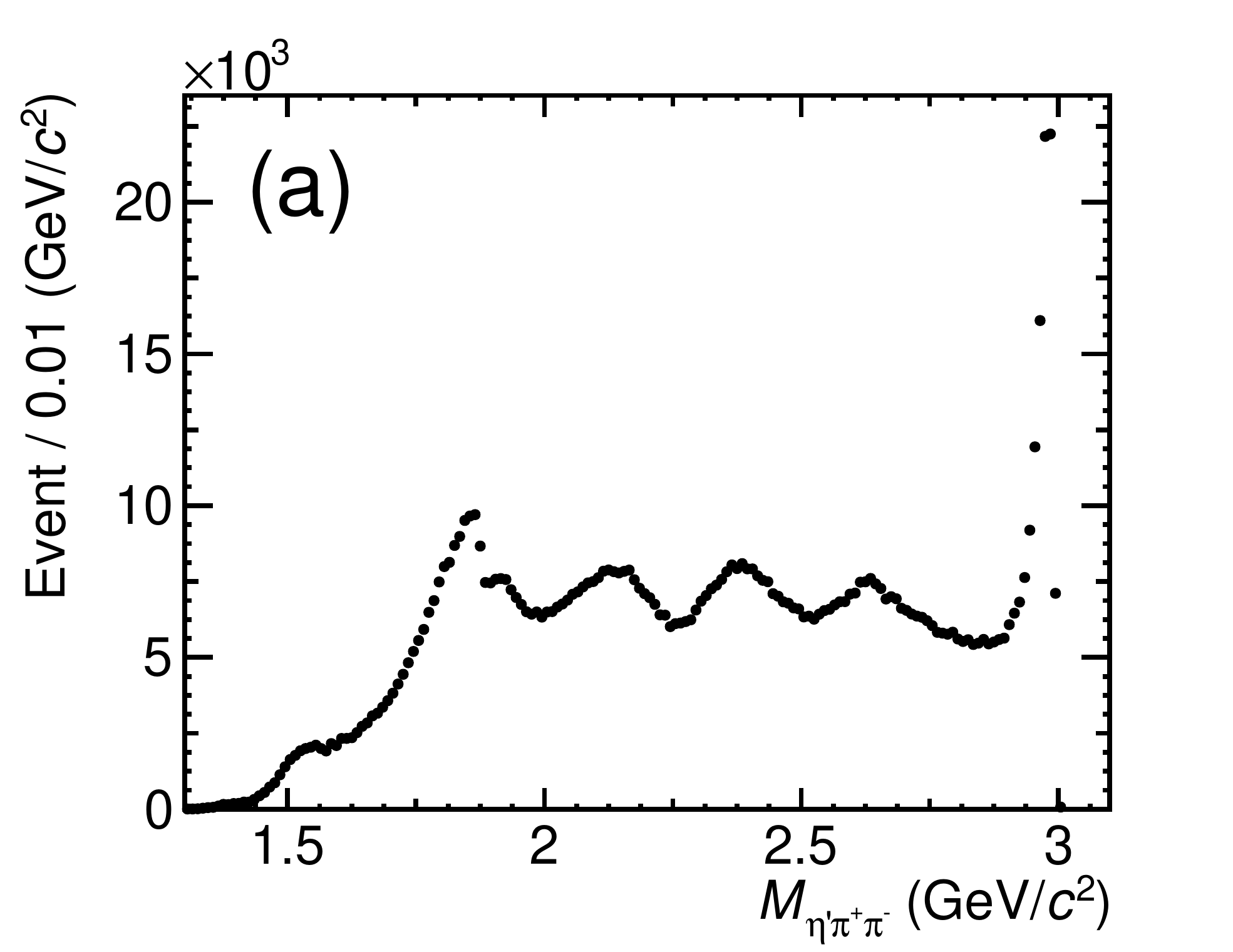}
  \label{fig:Subfigure1}
 }
 \subfloat{
  \includegraphics[width= 0.5\textwidth]{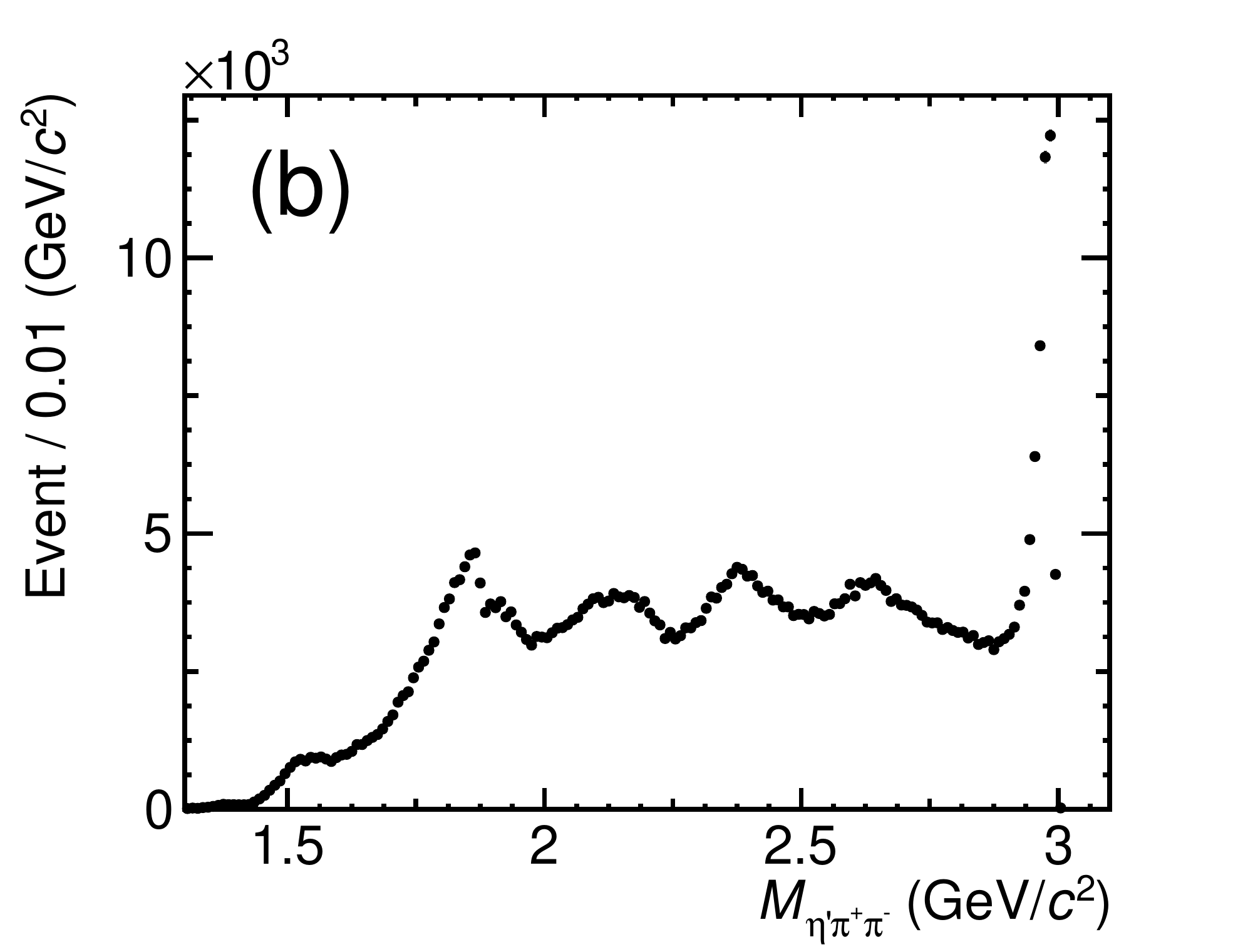}
  \label{fig:Subfigure2}
 }

 \subfloat{
  \includegraphics[width=0.5\textwidth]{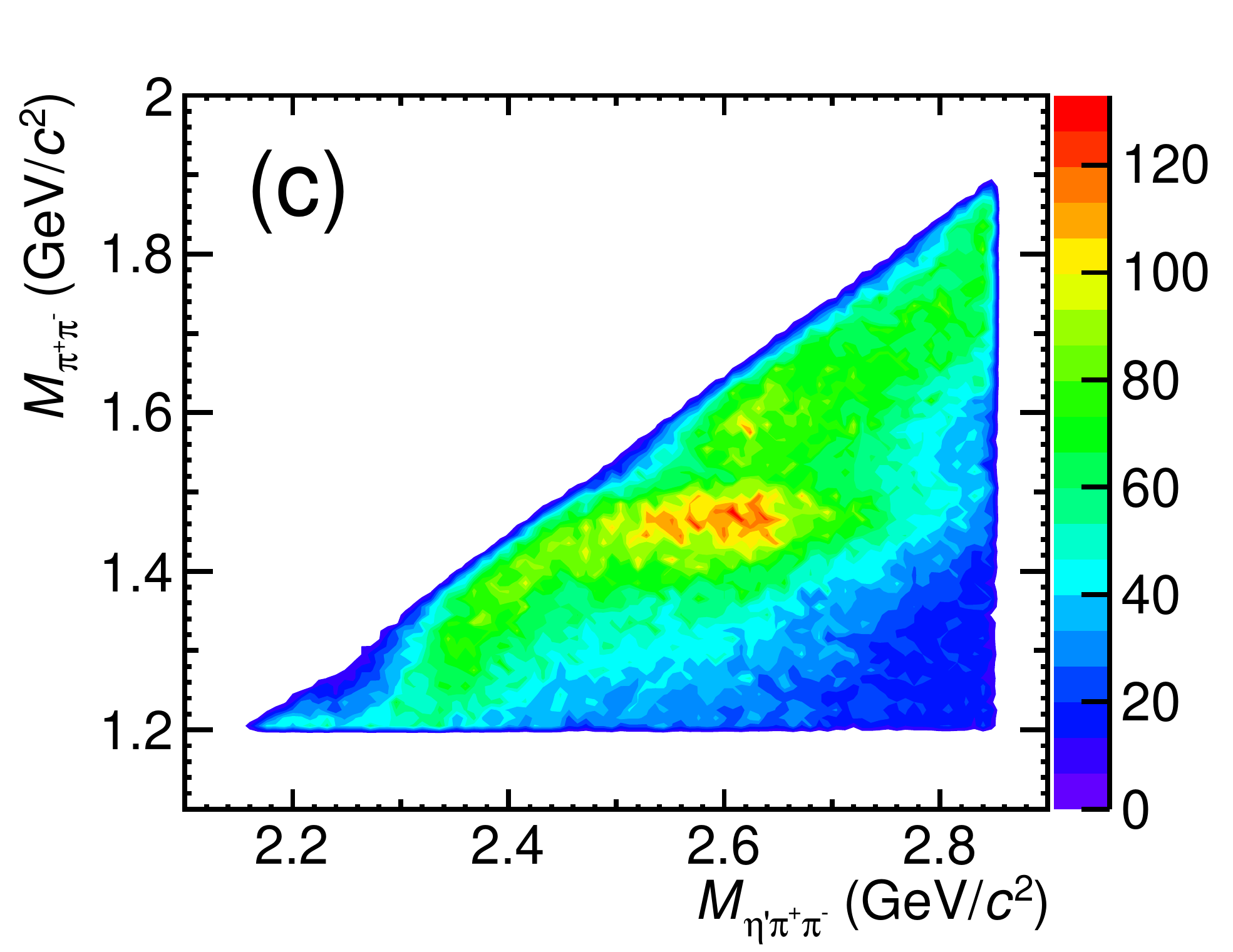}
  \label{fig:Subfigure3}
 }
 \subfloat{
  \includegraphics[width= 0.5\textwidth]{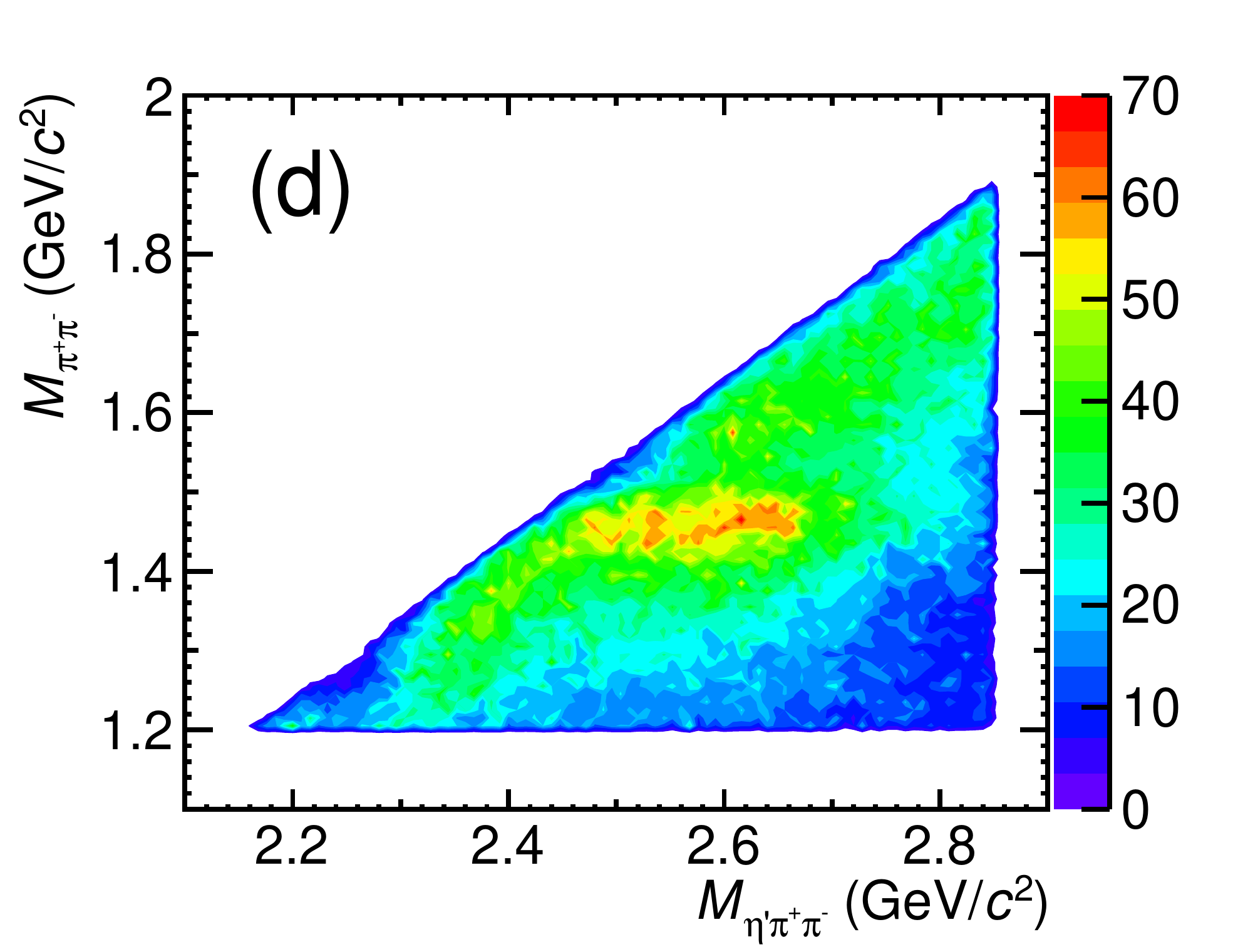}
  \label{fig:Subfigure4}
 }
 \caption{
  \protect The $J/\psi\rightarrow\gamma\pi^{+}\pi^{-}\eta'$ candidates in data: (top) the invariant mass spectrum of the final state $\pi^{+}\pi^{-}\eta'$ after event selection \textcolor{blue}{(a)} with the $\eta^{\prime}\rightarrow\gamma\pi^{+}\pi^{-}$ channel,
  \protect \textcolor{blue}{(b)} with the $\eta^{\prime}\rightarrow\pi^{+}\pi^{-}\eta$ channel, and (bottom) 
  \protect the two dimensional distribution of $M_{\pi^{+}\pi^{-}}$ versus $M_{\pi^{+}\pi^{-}\eta'}$ with $M_{\pi^{+}\pi^{-}} > 1.2~\mathrm{GeV}$/$c^{2}$ and $2.2 < M_{\pi^{+}\pi^{-}\eta^{\prime}} < 2.85~\mathrm{GeV}$/$c^{2}$ \textcolor{blue}{(c)} with the $\eta^{\prime}\rightarrow\gamma\pi^{+}\pi^{-}$ channel
  and \protect \textcolor{blue}{(d)} with the $\eta^{\prime}\rightarrow\pi^{+}\pi^{-}\eta$ channel.
 }
 \label{fig:fig1}
\end{figure}

For the $J/\psi\rightarrow\gamma\pi^{+}\pi^{-}\eta',
\eta'\rightarrow\gamma\pi^{+}\pi^{-}$ channel, event candidates are
required to have four charged tracks with at least three charged
tracks identified as pions and at least two photons with energies
larger than 100$~\mathrm{MeV}$.  A four-constraint (4C)
kinematic fit, which constrains the total four-momentum of all final
state particles to the initial four-momentum of the $e^{+}e^{-}$
system, is performed to the $\gamma\gamma\pi^{+}\pi^{-}\pi^{+}\pi^{-}$
hypothesis.  If there are more than two photons, the combination that
has the smallest $\chi^{2}_{\rm 4C}$ will be chosen, and
$\chi^{2}_{\rm 4C}<40$ is required.  The $\eta^{\prime}$ candidates
require $|M_{\gamma\pi^{+}\pi^{-}} - m_{\eta'}| <
15~\mathrm{MeV}/c^{2}$, where $m_{\eta'}$ is the mass of $\eta'$
reported by the PDG \cite{pdg}.  If there is more than one
$\gamma\pi^{+}\pi^{-}$ combination passing the above criteria, the
combination with minimum $|M_{\gamma\pi^{+}\pi^{-}} - m_{\eta'}|$ will
be selected.  For the photons of the selected combination, the requirements
$|M_{\gamma\gamma} - m_{\pi^{0}}|>40~\mathrm{MeV}/c^{2}$,
$|M_{\gamma\gamma} - m_{\eta}|>30~\mathrm{MeV}/c^{2}$, and
$720$ $<M_{\gamma\gamma}<820~\mathrm{MeV}/c^{2}$,
where $m_{\pi^{0}}$ and $m_{\eta}$ are the masses of $\pi^{0}$ and
$\eta$ mesons from the PDG \cite{pdg}, are used to suppress the
backgrounds from the processes of
$J/\psi\rightarrow\pi^{0}\pi^{+}\pi^{-}\pi^{+}\pi^{-}$,
$J/\psi\rightarrow\eta\pi^{+}\pi^{-}\pi^{+}\pi^{-}$, and
$J/\psi\rightarrow\omega(\omega\rightarrow\gamma\pi^{0})\pi^{+}\pi^{-}\pi^{+}\pi^{-}$.
The requirement of $400~\mathrm{MeV}/c^{2}$ $<M_{\gamma\pi^{+}\pi^{-}}
< 563~\mathrm{MeV}/c^{2}$ is used to suppress the background from the
processes of
$J/\psi\rightarrow\gamma\eta(\eta\rightarrow\gamma\pi^{+}\pi^{-})\pi^{+}\pi^{-}$
and
$J/\psi\rightarrow\gamma\eta(\eta\rightarrow\pi^{0}\pi^{+}\pi^{-})\pi^{+}\pi^{-}$.
In order to suppress the background from
$J/\psi\rightarrow\pi^{0}\pi^{+}\pi^{-}\eta'$ decays, the $J/\psi$
radiative photon is paired with all additional photons, and events
with any pairing with
$|M_{\gamma\gamma}-m_{\pi^{0}}|<15~\mathrm{MeV}/c^{2}$ are
rejected. After application of the above selection criteria, there are
evident signatures of the $X(1835)$, $X(2120)$, $X(2370)$, as well as
a distinct signal of the $\eta_c$ meson in the $\pi^{+}\pi^{-}\eta'$
invariant mass spectrum, shown in Fig.~\subref{fig:Subfigure1}, which
are consistent with previous BESIII results
\cite{X(1835)_confirmed}. In addition, there is a structure around
2.6$~\mathrm{GeV}$/$c^{2}$, the $X(2600)$, in the
$\pi^{+}\pi^{-}\eta'$ invariant mass spectrum, which is strongly
correlated to a structure around 1.5$~\mathrm{GeV}$/$c^{2}$ in the
$\pi^{+}\pi^{-}$ invariant mass spectrum as shown in
Fig.~\subref{fig:Subfigure3}.

For the $J/\psi\rightarrow\gamma\pi^{+}\pi^{-}\eta',
\eta'\rightarrow\pi^{+}\pi^{-}\eta,\ \eta\rightarrow\gamma\gamma$
channel, event candidates are required to have four charged tracks
with at least three charged tracks identified as pions and at least
three photons with energies larger than 100$~\mathrm{MeV}$.  A
4C kinematic fit is performed to the
$\gamma\gamma\gamma\pi^{+}\pi^{-}\pi^{+}\pi^{-}$ hypothesis. If there
are more than three photon candidates, the combination with the
smallest $\chi^{2}_{\rm 4C}$ will be chosen, and $\chi^{2}_{\rm
  4C}<40$ is required.  The $\eta$ candidates are reconstructed with
the requirement of $|M_{\gamma\gamma} -
m_{\eta}|<30~\mathrm{MeV}/c^{2}$. For the three selected photons, the
requirement $|M_{\gamma\gamma} - m_{\pi^{0}}|>40~\mathrm{MeV}/c^{2}$
is used for all photon pairs to suppress the $\pi^0$
background. Besides the 4C kinematic fit, a five-constraint (5C)
kinematic fit is performed, in which in addition to the constraint on
the total four momentum of the final-state particles, the invariant
mass of two photons coming from $\eta$ is constrained to $m_{\eta}$.
If more than one combination is found in an event, the combination
with the minimum $\chi^{2}_{\rm 5C}$ will be selected, and
$\chi^{2}_{\rm 5C}<40$ is required.  To select $\eta'$ candidates,
$|M_{\pi^{+}\pi^{-}\eta} - m_{\eta'}| < 10~\mathrm{MeV}/c^{2}$ is
required.  If there is more than one $\pi^{+}\pi^{-}\eta$ combination
passing the above criteria, the combination with smallest
$|M_{\pi^{+}\pi^{-}\eta} - m_{\eta'}|$ will be selected as the $\eta'$
candidate. In order to suppress the background from the processes of
$J/\psi\rightarrow\pi^{0}\pi^{+}\pi^{-}\eta'$, the $J/\psi$ radiative
photon is paired with all additional photons, and events with any pair
with $|M_{\gamma\gamma} - m_{\pi^{0}}|<15~\mathrm{MeV}/c^{2}$ are
rejected. After the above selection criteria, the
$\pi^{+}\pi^{-}\eta'$ mass spectrum as shown in
Fig.~\subref{fig:Subfigure2} is similar to that in the
$\eta^{\prime}\rightarrow\gamma\pi^{+}\pi^{-}$ channel. There is a
structure around 2.6$~\mathrm{GeV}$/$c^{2}$, the $X(2600)$, in the
$\pi^{+}\pi^{-}\eta^{\prime}$ invariant mass spectrum, which is
strongly correlated with the structure around
1.5$~\mathrm{GeV}$/$c^{2}$ in the $\pi^{+}\pi^{-}$ invariant mass
spectrum as shown in Fig.~\subref{fig:Subfigure4}.

Possible background contributions are studied using an inclusive MC
sample. There are two kinds of background. One is from
non-$\eta^{\prime}$ processes and the other is from the process
$J/\psi\rightarrow\pi^{0}\pi^{+}\pi^{-}\eta'$. The former one can be
estimated with the $\eta^{\prime}$ side-band regions in data, which
are chosen to be $30 < |M_{\gamma\pi^{+}\pi^{-}} - m_{\eta'}| <
45~\mathrm{MeV}/c^{2}$ for the channel of
$\eta'\rightarrow\gamma\pi^{+}\pi^{-}$, and $20 <
|{\it{M}}_{\pi^{+}\pi^{-}\eta} - m_{\eta'}| < 30~\mathrm{MeV}/c^{2}$
for the $\eta'\rightarrow\pi^{+}\pi^{-}\eta,\
\eta\rightarrow\gamma\gamma$ channel.  Background coming from
$J/\psi\rightarrow\pi^{0}\pi^{+}\pi^{-}\eta'$ decays can pass the
final selection criteria for
$J/\psi\rightarrow\gamma\pi^{+}\pi^{-}\eta'$ decays if one of the
photons from the $\pi^{0}$ decay is not reconstructed or is out of the
detector acceptance. To estimate the background contribution from the
$J/\psi\rightarrow\pi^{0}\pi^{+}\pi^{-}\eta'$ decays, we use a control
sample of decays passing the selection criteria for
$J/\psi\rightarrow\gamma\pi^{+}\pi^{-}\eta'$ decays, but with a
reversed $\pi^0$ veto criterium,
$|M_{\gamma\gamma}-m_{\pi^{0}}|<15~\mathrm{MeV}/c^{2}$.  The residual
$\pi^{0}\pi^{+}\pi^{-}\eta'$ background contribution in the
$J/\psi\rightarrow\gamma\pi^{+}\pi^{-}\eta'$ signal region can be
estimated by reweighting the events from the control sample. The
weight factors are dependent on the radiative photon energy, and equal
to the MC efficiency ratio of the
$J/\psi\rightarrow\gamma\pi^{+}\pi^{-}\eta'$ signal selection and the
$J/\psi\rightarrow\pi^{0}\pi^{+}\pi^{-}\eta'$ background sample
selection. Neither of these background components produces a peaking
structure in the $\pi^{+}\pi^{-}\eta'$ and $\pi^{+}\pi^{-}$ invariant
mass spectrum.

In order to determine the signal of the $X(2600)$ resonance with a
consequent decay to a resonance at mass around
1.5$~\mathrm{GeV}$/$c^{2}$ in the $\pi^{+}\pi^{-}$ invariant mass
spectrum, a simultaneous fit to the $\pi^{+}\pi^{-}\eta'$ and
$\pi^{+}\pi^{-}$ mass spectra is performed, including the two decay
channels of $\eta'\rightarrow\gamma\pi^{+}\pi^{-}$ and
$\eta'\rightarrow\pi^{+}\pi^{-}\eta,\ \eta\rightarrow\gamma\gamma$. In
the fit, the number of events is the same in the two projected mass
spectra for a given channel. Moreover, the mass, width and branching
fraction of each resonance are common between the two $\eta'$ decay
channels in the simultaneous fit. The line shape of the $X(2600)$
resonance in the $\pi^{+}\pi^{-}\eta'$ mass spectrum is described with
an efficiency-corrected Breit-Wigner function convolved with a double
Gaussian function describing the detector resolution.  The FWHM (full
width at half maximum) resolution on the $M_{\pi^{+}\pi^{-}\eta'}$
distribution is around 10$~\mathrm{MeV}$/$c^{2}$.  The structure
around 1.5$~\mathrm{GeV}$/$c^{2}$ in the $\pi^+\pi^-$ mass spectrum is
described with an efficiency-corrected interference between the
$f_{0}(1500)$ and the $f_{2}^{\prime}(1525)$ convolved with a double
Gaussian function describing the detector resolution.  The
$M_{\pi^{+}\pi^{-}}$ FWHM resolution is about
9$~\mathrm{MeV}$/$c^{2}$.  Two Breit-Wigner functions are used to
describe the line shape of $f_{0}(1500)$ and
$f_{2}^{\prime}(1525)$. The mass and width of the
$f_{2}^{\prime}(1525)$ are fixed to the PDG values \cite{pdg}, while
the mass and width of the $f_{0}(1500)$ are floated in the fit, given
the large discrepancy between different experiments
\cite{besII_1500,besIII_1500,Aaij:2014emv,Bertin:1998hu,Aubert:2006nu}.
The contributions from other processes with the
$\gamma\pi^{+}\pi^{-}\eta'$ final state are described with two
different fourth order polynomial functions in the $\pi^{+}\pi^{-}\eta'$
and $\pi^{+}\pi^{-}$ mass spectra, and are treated as incoherent.
The background contributions from the non-$\eta^{\prime}$ events and
$J/\psi\rightarrow\pi^0\pi^{+}\pi^{-}\eta'$ decays are estimated with the
two different methods as described earlier, and both the
mass line shapes and yields are fixed in the fit.

\begin{figure}[htbp]
	\centering
	\subfloat{
		\includegraphics[width=0.5\textwidth]{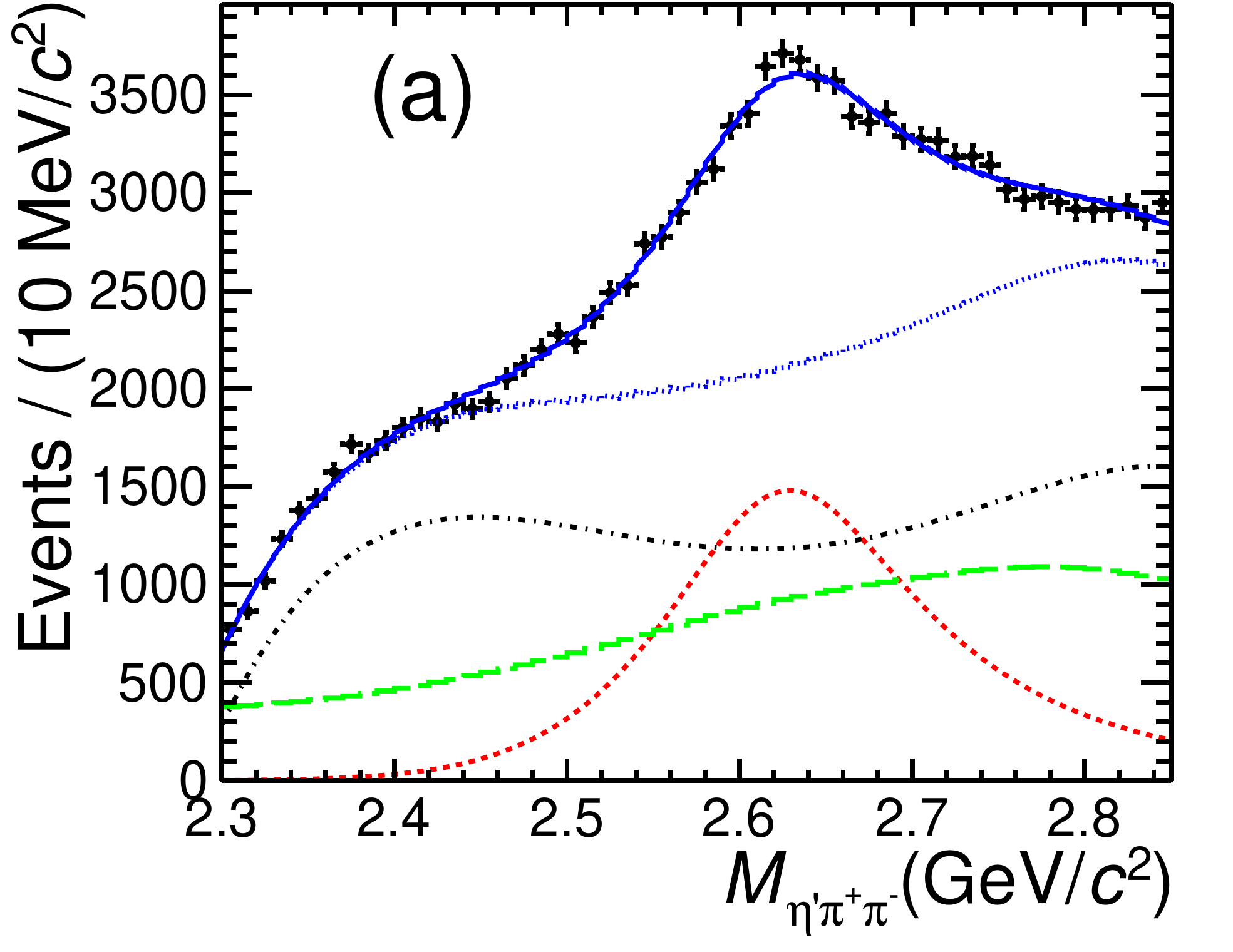}
		\label{fig:Subfigure5}
	}
	\subfloat{
		\includegraphics[width=0.5\textwidth]{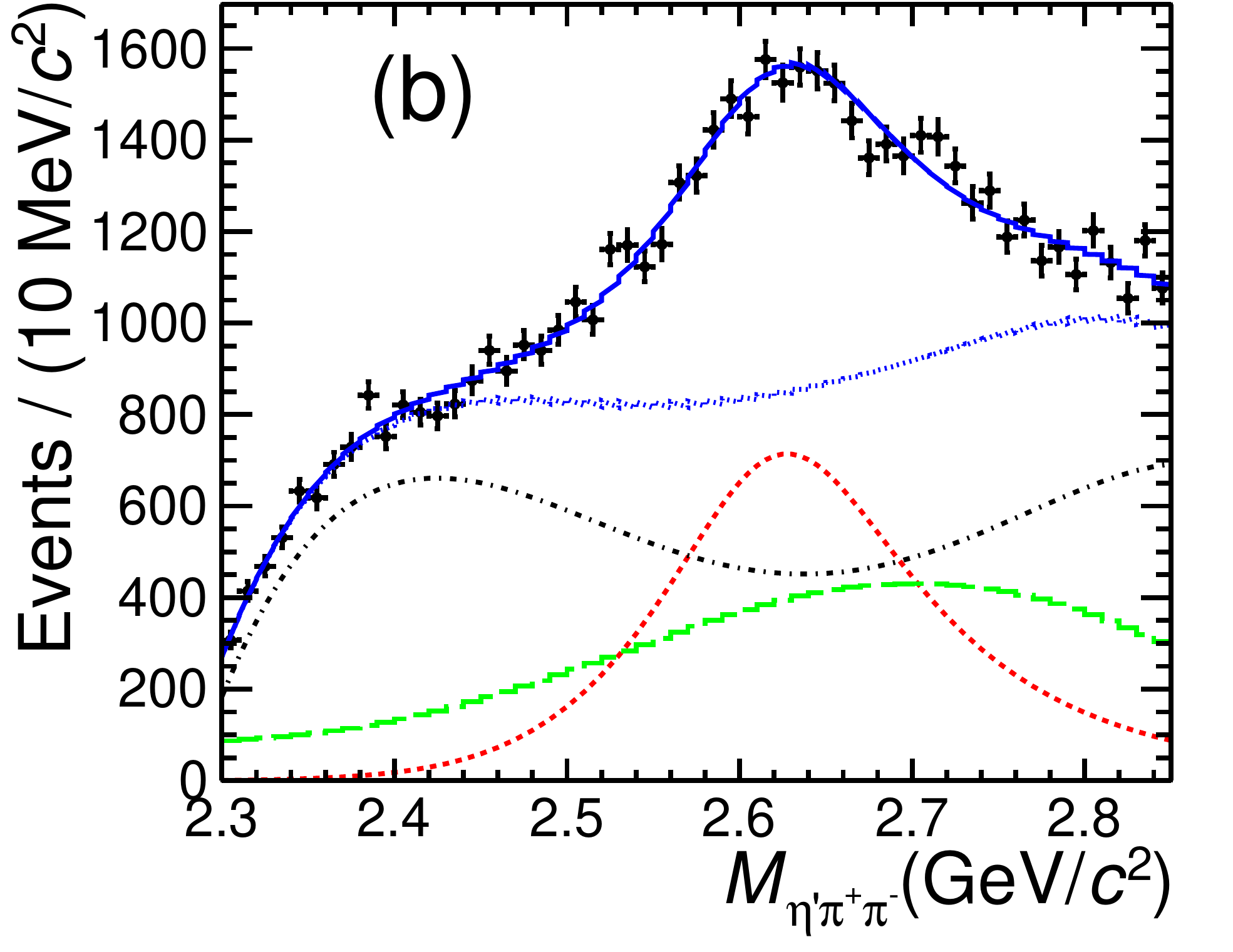}
		\label{fig:Subfigure6}
	}

	\subfloat{
		 \includegraphics[width=0.5\textwidth]{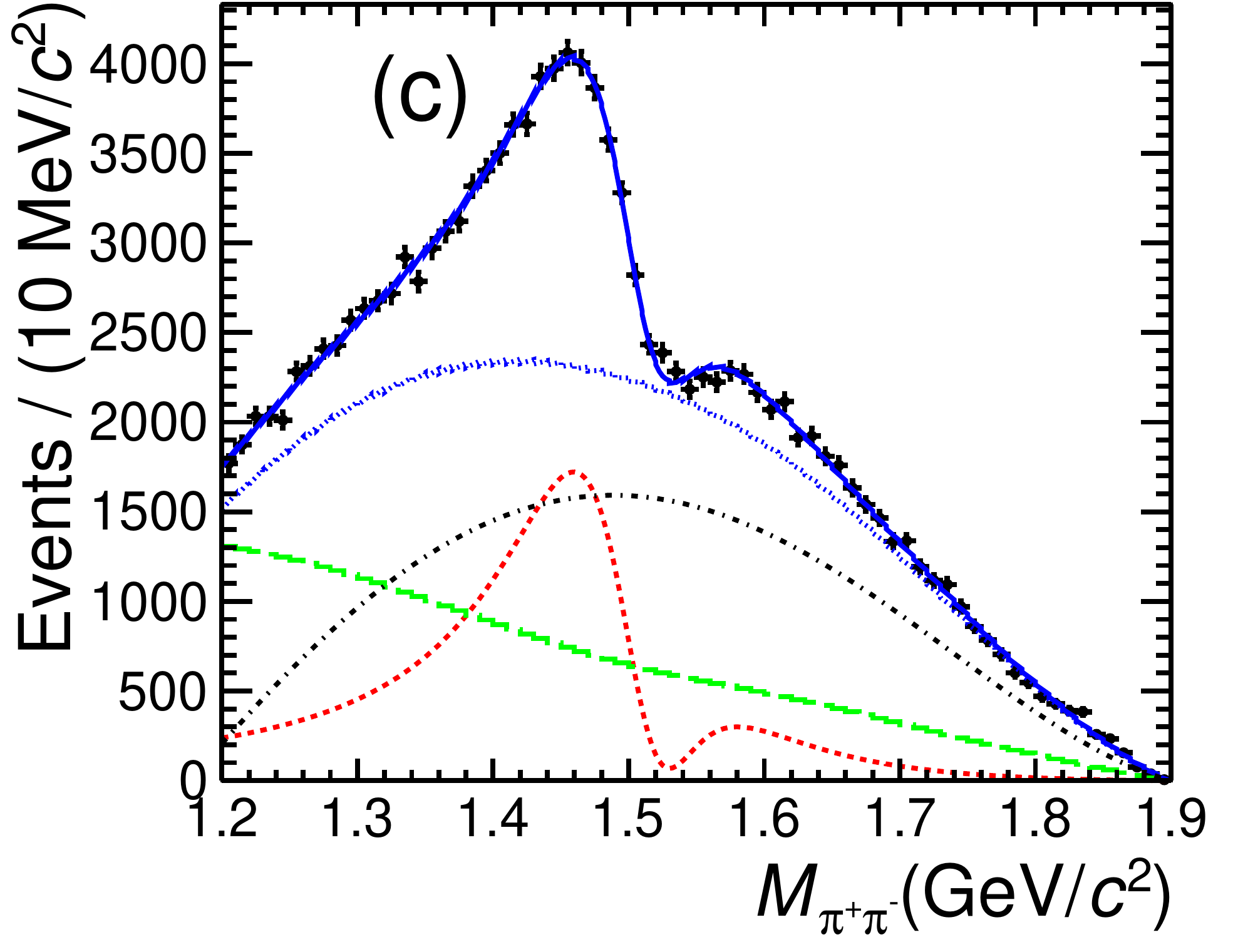}
		 \label{fig:Subfigure7}
	}
	\subfloat{
		 \includegraphics[width=0.5\textwidth]{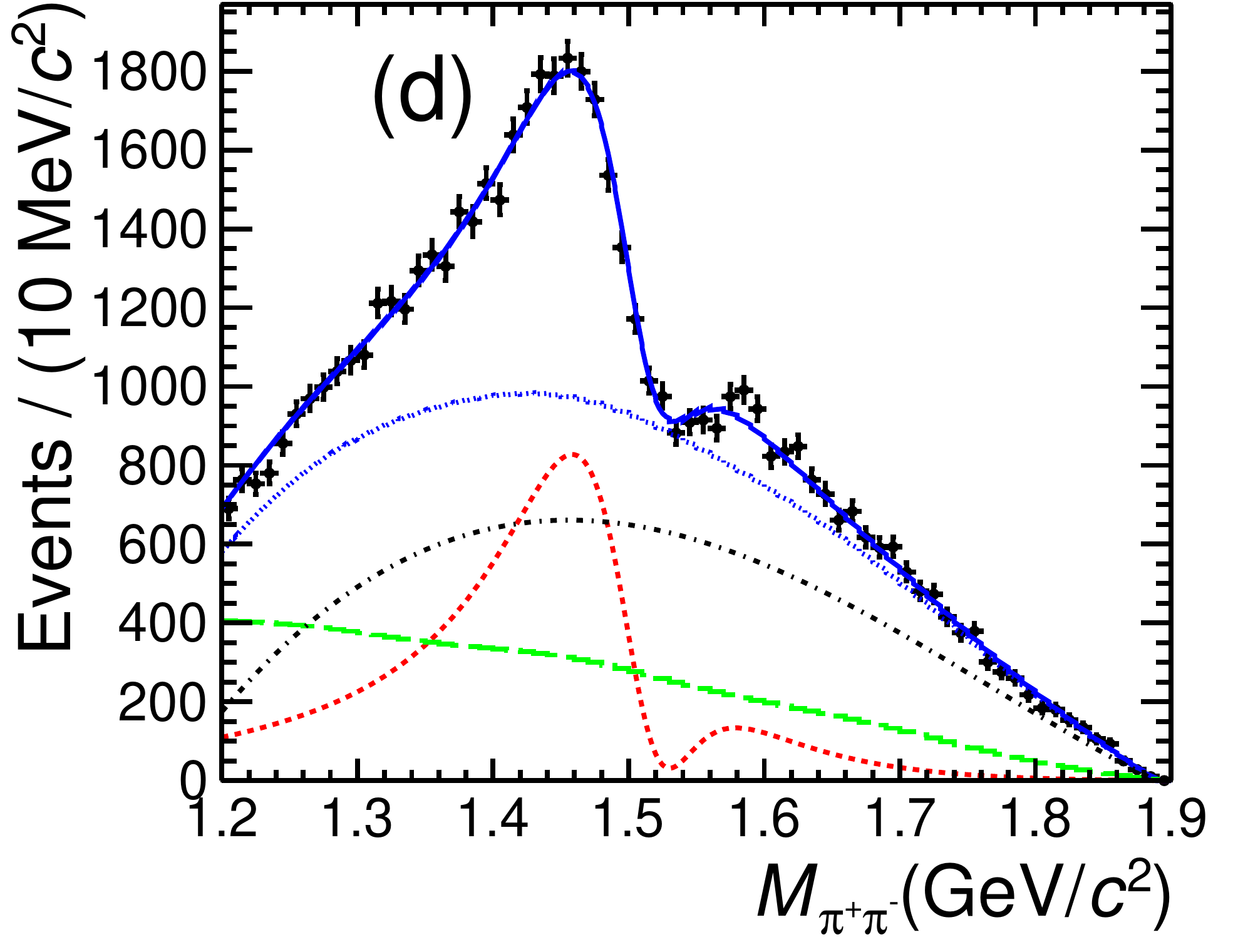}
		 \label{fig:Subfigure8}
	}
	\caption{The $\pi^{+}\pi^{-}\eta'$ and $\pi^{+}\pi^{-}$ mass spectra distributions with the two decay channels of $\eta'$, $\eta'\rightarrow\gamma\pi^{+}\pi^{-}$ and $\eta'\rightarrow\pi^{+}\pi^{-}\eta$, with the simultaneous fit results overlaid: \protect \textcolor{blue}{(a)} and \protect\textcolor{blue}{(c)} are fit results for $\eta'\rightarrow\gamma\pi^{+}\pi^{-}$ channel, \protect \textcolor{blue}{(b)} and \protect\textcolor{blue}{(d)} are fit results for the $\eta'\rightarrow\pi^{+}\pi^{-}\eta$ channel. The dots with error bar are data, the blue solid lines are the total fits, the red dashed lines describe the $X(2600)$ signal in the $\pi^{+}\pi^{-}\eta'$ mass spectrum, and the structure around 1.5$~\mathrm{GeV}$/$c^{2}$ in $\pi^{+}\pi^{-}$ mass spectrum, the black dash-dotted lines correspond to the background described with a polynomial function, and the green long dashed lines are	$J/\psi\rightarrow\pi^{0}\pi^{+}\pi^{-}\eta'$ and non-$\eta'$ background, the blue dotted lines are the total background, including the $J/\psi\rightarrow\pi^{0}\pi^{+}\pi^{-}\eta'$, non-$\eta'$ background and polynomial background.}
	\label{fig:fig5}
\end{figure}

Figure~\ref{fig:fig5} shows the simultaneous fit results with the two
$\eta^\prime$ decay modes in the region with
$M_{\pi^{+}\pi^{-}} > 1.2~\mathrm{GeV}$/$c^{2}$ and $2.3 <
M_{\pi^{+}\pi^{-}\eta'} < 2.85~\mathrm{GeV}$/$c^{2}$.  The statistical
significance is determined from the change of $-2\ln{\it{L}}$
(${\it{L}}$ is the combined likelihood of simultaneous fit) in the fit
with and without signal assumption, considering the change of degrees
of freedom of the fits. The significances of the $X(2600)$,
$f_{0}(1500)$ and $f_{2}^{\prime}(1525)$ resonances are all larger
than 20$\sigma$. The mass and width of the $X(2600)$ are
$2617.8\pm2.1$(stat.)$~\mathrm{MeV}$/$c^{2}$ and $200\pm8$(stat.)$~\mathrm{MeV}$, respectively, and
those of the $f_{0}(1500)$ are $1498.0\pm4.5$(stat.)$~\mathrm{MeV}$/$c^{2}$ and
$166\pm10$(stat.)$~\mathrm{MeV}$, respectively.  The masses and widths
are summarized in Table~\ref{mass&widths}.
The branching fractions 
are measured to be $B(J/\psi\rightarrow\gamma X(2600))\cdot
B(X(2600)\rightarrow f_{0}(1500)\eta')\cdot
B(f_{0}(1500)\rightarrow\pi^{+}\pi^{-})$ =
($3.39\pm0.18$(stat.))$\times10^{-5}$ and $B(J/\psi\rightarrow\gamma
X(2600))\cdot B(X(2600)\rightarrow f_{2}^{'}(1525)\eta')\cdot
B(f_{2}^{'}(1525)\rightarrow\pi^{+}\pi^{-})$ =
($2.43\pm0.13$(stat.))$\times10^{-5}$.  
The numbers of signal events, efficiencies, and branching fractions are
listed in Table~\ref{BFs}. 

The consistency between the two $\eta'$ decay channels is verified by
fitting the two channels separately with the method
described above.  The fit to the
$\eta'\rightarrow\gamma\pi^{+}\pi^{-}$ channel gives $M =
2619.5\pm2.3$(stat.)$~\mathrm{MeV}$/$c^{2}$ and
$\Gamma=185\pm10$(stat.)$~\mathrm{MeV}$. 
For the fit to the $\eta'\rightarrow\pi^{+}\pi^{-}\eta$ channel, the
mass and width of $X(2600)$ are determined to be $M =
2616.1\pm3.9$(stat.)$~\mathrm{MeV}$/$c^{2}$ and
$\Gamma=206\pm13$(stat.)$~\mathrm{MeV}$. 
The statistical significances of the two $\eta'$ decay
channels are both greater than 10$\sigma$.  The masses and widths of the
$X(2600)$ are in good agreement between the two $\eta'$ decay
channels.

The systematic uncertainties of the mass and width are mainly from the
fit strategy, fit range, background estimation, phase space of
$J/\psi\rightarrow\gamma\pi^{+}\pi^{-}\eta'$ decays, line shape of the
intermediate resonance $f^{'}_{2}(1525)$, quantum number assumption of
the $X(2600)$ state and treatment of the contribution from the other
processes with the final state of $\gamma\pi^{+}\pi^{-}\eta'$.  The
uncertainties are determined by including an additional resonance in
the fit, using different parameters for the $f^{'}_{2}(1525)$ line
shape, a different quantum number assignment for the $X(2600)$, and
different treatment of the contribution from the other processes, as
well as varying the fit range, side-band region, etc..  The dominant
uncertainty sources are from including an additional resonance in the
fit and assuming different quantum numbers for the $X(2600)$.  To
estimate the uncertainty from the fit with the additional resonance,
the $X(2370)$ is included into the simultaneous fit, and
interference between the $X(2370)$ and the $X(2600)$ is included.  Two
different hypotheses for the quantum numbers of $X(2600)$, $0^{-+}$
and $2^{-+}$, are used in the simultaneous fit, the difference
between them is treated as the uncertainty caused by the quantum
number assumption for the $X(2600)$ state.  The largest deviations
from the baseline case are taken as the systematic errors. The total
uncertainties on the mass and width of the $X(2600)$ are
$^{+18.2}_{-1.9}~\mathrm{MeV}/c^{2}$ and $^{+20}_{-17}~\mathrm{MeV}$.

To estimate the systematic error of the branching fraction
measurement, additional uncertainties are considered, including the
data-MC difference in the charged track reconstruction efficiency,
photon detection efficiency, PID efficiency, the
kinematic fit and the number of $J/\psi$ events.  The total
systematic uncertainties on the branching fraction are
$^{+27\%}_{-20\%}$ and $^{+13\%}_{-47\%}$ for the
$B(J/\psi\rightarrow\gamma X(2600))\cdot B(X(2600)\rightarrow
f_{0}(1500)\eta')\cdot B(f_{0}(1500)\rightarrow\pi^{+}\pi^{-})$ and
$B(J/\psi\rightarrow\gamma X(2600))\cdot B(X(2600)\rightarrow
f_{2}^{'}(1525)\eta')\cdot
B(f_{2}^{'}(1525)\rightarrow\pi^{+}\pi^{-})$, respectively.

\begin{table}[bth]
\caption{Masses and widths of the $ f_0(1500)$ and
  $X(2600)$.  The first uncertainties are statistical, and the second
  are systematic. \label{mass&widths}}
\begin{tabular}{lcc}
\hline
\T Case & $ f_0(1500)$ & $X(2600)$ \B \\\hline
\T Mass (MeV/$c^2$)    & $1498.0 \pm 4.5^{+4.0}_{-15.2}$ & $2617.8 \pm 2.1^{+18.2}_{-1.9}$\\
\T Width (MeV)         & $166 \pm 10^{+13}_{-26} $      & $200 \pm
8^{+20}_{-17}$ \B \\\hline
\end{tabular}
\end{table}

\begin{table}[bth]
\caption{Branching fractions for $ J/\psi \to \gamma X(2600),~X(2600)
  \to f_0(1500)/f_{2}^{'}(1525)\eta',~f_0(1500)/f_{2}^{'}(1525)\to \pi^+\pi^-$. BF is the
  product of the three branching fractions involved. The first
  uncertainties are statistical, and the second are
  systematic. \label{BFs}}
\begin{tabular}{lcc}
\hline
\T Case        &  $f_0(1500)$    & $f_{2}^{'}(1525)$  \B \\ \hline
\T Events      &  $26917 \pm 1069$  & $19094 \pm759$   \\
Efficiency  &    19\%          &     15\%            \\
\T BF ($\times 10^{-5}$) & $3.39\pm0.18^{+0.91}_{-0.66}$ &
  $2.43\pm0.13^{+0.31}_{-1.11}$ \B \\ \hline
\end{tabular}
\end{table}

In summary, the process $J/\psi\rightarrow\gamma\pi^{+}\pi^{-}\eta'$,
with $\eta'\rightarrow\gamma\pi^{+}\pi^{-}$ and
$\eta'\rightarrow\pi^{+}\pi^{-}\eta,\ \eta\rightarrow\gamma\gamma$, is
studied with $(10087\pm44)\times10^{6}$ $J/\psi$ events collected by
the BESIII detector. A resonance, the $X(2600)$, is
observed for the first time, with a statistical significance greater
than 20 $\sigma$.  There is a strong correlation between the $X(2600)$
and the structure at 1.5$~\mathrm{GeV}$/$c^{2}$ in the
$\pi^{+}\pi^{-}$ invariant mass spectrum.  A simultaneous fit of the
$\pi^{+}\pi^{-}\eta'$ and $\pi^{+}\pi^{-}$ mass spectra with the two
$\eta'$ decay channels of $\eta'\rightarrow\gamma\pi^{+}\pi^{-}$ and
$\eta'\rightarrow\pi^{+}\pi^{-}\eta$ is performed.
The structure around 1.5$~\mathrm{GeV}$/$c^{2}$ in the
$\pi^{+}\pi^{-}$ invariant mass spectra can be well described with the
interference between the $f_{0}(1500)$ and the $f_{2}^{'}(1525)$
resonances.  The masses and width of the $X(2600)$ and $f_0(1500)$
resonances are listed in Table~\ref{mass&widths}, and the product
branching fractions $B(J/\psi\rightarrow\gamma X(2600))\cdot
B(X(2600)\rightarrow f_{0}(1500)/f_{2}^{'}(1525)\eta')\cdot
B(f_{0}(1500)/f_{2}^{'}(1525)\rightarrow\pi^{+}\pi^{-})$ are listed in
Table~\ref{BFs}.

With these decay modes, the spin parity assignment of $0^{-+}$ or
$2^{-+}$ is favored for the $X(2600)$.  In order to understand the
nature of the $X(2600)$ state, whether it can be interpreted as an
$\eta$ radial excitation \cite{PhysRevD.102.114034}, or a potential
exotic hadron, it is important to determine its spin-parity and to
study its production and decay properties in other $J/\psi$ decay
channels.

\vspace{0.2cm}
The BESIII collaboration thanks the staff of BEPCII and the IHEP computing center for their strong support. 
This work is supported in part by National Key Research and Development Program of China under Contracts Nos. 2020YFA0406300, 2020YFA0406400; 
National Natural Science Foundation of China (NSFC) under Contracts Nos. 11625523, 11635010, 11735014, 11822506, 11835012, 11935015, 11935016, 11935018, 11961141012, 12022510, 12025502, 12035009, 12035013, 12061131003; 
the Chinese Academy of Sciences (CAS) Large-Scale Scientific Facility Program; 
Joint Large-Scale Scientific Facility Funds of the NSFC and CAS under Contracts Nos. U1732263, U1832207; 
CAS Key Research Program of Frontier Sciences under Contract No. QYZDJ-SSW-SLH040; 100 Talents Program of CAS; 
INPAC and Shanghai Key Laboratory for Particle Physics and Cosmology; ERC under Contract No. 758462; 
European Union Horizon 2020 research and innovation programme under Contract No. Marie Sklodowska-Curie grant agreement No 894790; 
German Research Foundation DFG under Contracts Nos. 443159800, Collaborative Research Center CRC 1044, GRK 214; 
Istituto Nazionale di Fisica Nucleare, Italy; Ministry of Development of Turkey under Contract No. DPT2006K-120470; 
National Science and Technology fund; Olle Engkvist Foundation under Contract No. 200-0605; STFC (United Kingdom); 
The Knut and Alice Wallenberg Foundation (Sweden) under Contract No. 2016.0157; The Royal Society, UK under Contracts Nos. DH140054, DH160214; 
The Swedish Research Council; U. S. Department of Energy under Contracts Nos. DE-FG02-05ER41374, DE-SC-0012069.
\nocite{*}

\bibliographystyle{apsrev4-1}
\bibliography{draft}

\end{document}